# Survey of ground state neutron spectroscopic factors from Li to Cr isotopes

M.B. Tsang[1*], Jenny Lee[1,2], W.G. Lynch[1F]

[1]*National Superconducting Cyclotron Laboratory and Department of Physics and Astronomy, Michigan State University, East Lansing, MI 48824*
[2]*Physics Department, Chinese University of Hong Kong, Shatin, Hong Kong, China*

## Abstract

The ground state neutron spectroscopic factors for 80 nuclei ranging in Z from 3 to 24 have been extracted by analyzing the past measurements of the angular distributions from (d,p) and (p,d) reactions. We demonstrate an approach that provides systematic and consistent values with minimum assumptions. For the 61 nuclei that have been described by large-basis shell-model calculations, most experimental spectroscopic factors are reproduced to within 20%.

[*] Corresponding author: tsang@nscl.msu.edu

Present and planned rare isotope accelerators offer opportunities to explore the structure of unstable nuclei. Like their stable counterparts, the structure of these nuclei reflects interplay between single-particle degrees of freedom that govern the shell structure of nuclei [1] and collective degrees of freedom that govern nuclear deformations [2]. Single-particle properties, such as the shell closure at "magic" neutron or proton numbers, can be approximated by the independent particle model, which assumes that nucleons move in a mean-field potential [1]. Residual nucleon-nucleon interactions, however, mix the single-particle states of the independent particle model [3-5]. Understanding this mixing is an important objective of modern structure calculations; it alters the occupancies, or spectroscopic factors (SF), of single-nucleon orbits from their independent particle model values, and it can even modify the "magic" numbers of nuclei that have a large neutron excess or are near the proton drip line [6].

Experimental measurements of the SF's for single-nucleon orbits within the quantum states of nuclei provide the key experimental probe of single-particle dynamics. Both transfer and knockout reactions have been widely studied in order to determine the SF's of single-nucleon orbits. Transfer reactions comprise the preponderance of such studies in the past four decades. However, their extracted SF's often varied widely, reflecting inconsistencies in the choice of optical potentials for the incoming and outgoing channels to which the transfer cross sections are sensitive [7,8]. In this paper, we reanalyze most of the (p,d) and (d,p) neutron transfer reactions between 10 and 60 MeV incident energy and demonstrate that the use of global optical potentials, not available for the earlier studies, allows systematic extraction of SF's that are remarkably consistent with large-basis shell-model (LB-SM) calculations [9]. With some modifications, these procedures can be readily extended to rare isotope beam experiments, because they do not require the availability of elastic scattering data for the systems under investigation.

In the present work, we restrict our analysis to ground-state-to-ground-state transitions of A(p,d)B and the inverse B(d,p)A reactions. Both reactions probe the angular momentum, the orbital occupancy, and wavefunction of the valence neutron in the nuclear surface. The overlap integral between the wave function of one state in nucleus A and another in B defines the theoretical neutron SF for transfer between these states. The ratio of the measured cross-section divided by the cross section calculated with a reaction model provides its experimental counterpart [3-5].

In the Distorted Born Wave Approximation (DWBA), transfer reactions are described as a fast one-step process [3-5]. In the present work, the approach is not strictly the DWBA because we adopt the deuteron optical potential calculated using the Johnson-Soper adiabatic model, which corrects approximately for deuteron breakup in the mean field of the target [10]. Following ref. [8], the Chapel-Hill 89 [11] global nucleon-nucleus optical potentials are used and these neutron and proton potentials are folded to obtain the deuteron optical potential. Nonlocality corrections with range parameters of 0.85 fm and 0.54 fm for the proton and deuteron channels, respectively are assumed [12]. The deuteron finite range corrections to the DWBA integral are calculated using the local energy approximation and the strength ($D_o^2$=150006.25 fm$^3$) and range ($\beta$=0.7457 fm) parameters of the Reid soft-core $^3S_1$-$^3D_1$ neutron-proton interaction [13]. For simplicity, a central neutron potential of Woods-Saxon shape with fixed radius ($r_o$=1.25 fm) and diffuseness ($a_o$=0.65 fm) parameters is assumed and the depth of potential is adjusted to reproduce the experimental binding energy. We use the University of Surrey version of TWOFNR [14], a direct reaction model code, to calculate the angular distributions. Other widely used reaction model codes, DWUCK5 [15] and FRESCO [16], yield nearly identical predictions with the same input information [17-19].

Table 1 lists the experimental and theoretical SF values determined for 80 nuclei, from $^6$Li to $^{55}$Cr, studied in this work. The range of neutron separation energies range from 0.5 to 19 MeV. The SF values range from very small (<0.1 for $^{20}$F, $^{21}$Ne, $^{47}$Ti) to rather large (~7 for $^{48}$Ca). In the interest of brevity, details of our evaluation of the data will be discussed in another paper [20, 21].

The uncertainties associated with the SF's listed in Table 1 are determined by comparing SF values extracted from the pickup (p,d) reaction with those extracted from the inverse stripping (d,p) reaction corresponding to the same ground state valence neutron. Table II lists the nuclei studied with both types of reactions. The averaged SF values obtained from the (d,p) and (p,d) reactions are listed in the 2$^{nd}$ and 4$^{th}$ columns, respectively, and the corresponding numbers of measurements contributing to these averages are listed in the 3rd and 5th columns. As expected, there appears to be no systematic difference between the SF's determined by the (p,d) and (d,p) reactions. The scatter of the data points can be used to assess random uncertainties in the procedure and in the quality of the data. Assuming the fractional random uncertainty of each of the SF values to be the same, we find the data to be consistent with a 20% random uncertainty in each measurement. Comparisons of repeated measurements of the same reaction at the same energy suggest that experimental uncertainties may be the dominant contribution to this random uncertainty [8, 18, 20]. The uncertainties associated with our SF values are much less than that deduced by Endt (50%) [22] who "averaged" the (p,d) and (d,p) SF values obtained by various authors. The 50% uncertainty assigned by Endt reflected inconsistencies in the (p,d) and (d,p) analyses of different authors [8]; to obtain his "best SF values" with uncertainties of 25%, Endt averaged results from (p,d), (d,p), (d,t), and ($^3$He,$\alpha$) reactions [22].

Spectroscopic factors for nuclei with even number of valence neutrons generally exceed those of the neighboring nuclei with odd number of valence neutrons. This results from the pairing interaction, which couples pairs of neutrons to spin zero similar to the Cooper pairs in a superconductor. For nuclei in the vicinity of a closed shell, this trend can be well replicated by calculations that consider only pairing modifications to the independent particle model [IPM]. Assuming maximal pairing (minimum Seniority), one can obtain a simple relationship between the spectroscopic factor and the number of valence nucleons *(n)* with total angular momentum *j* [3]

$$\text{SF}= n \quad \text{for n=even}; \qquad \text{SF}=1-\frac{n-1}{2j+1} \quad \text{for n=odd} \qquad (1)$$

The thin bars in Figure 1 show the predictions of Eq. 1 as a function of the mass number A for the transfer of an $f_{7/2}$ neutron to or from Ca isotopes; the extracted neutron SF's are represented by star symbols. The excellent agreement reflects the fact that configuration mixing of $f_{7/2}$ neutrons outside the double magic $^{40}$Ca core is well described by a pairing interaction, with little discernable contribution from core polarization or higher lying orbits.

Most nuclei display more significant configuration mixing than these Ca isotopes. Figure 2 compares experimentally extracted SF's for 49 nuclei with predictions from two

models – the independent particle model plus pairing (left panel) and the LB-SM (right panel). Open symbols represent odd Z elements and the closed symbols represent even Z elements. The solid line indicates perfect agreement. Most extracted values are less than the predictions of the IPM-plus-maximal-pairing as represented by Eq. 1. If all available SF values are included in the left panel, the observed scatter of the data remains about the same.

If one diagonalizes the residual interaction within a LB-SM [6], that involves the mixing of several different orbitals, one can obtain a better description of nuclei. Using Oxbash [23] and the PPN, SPSDPF, SDPN and FPPN interactions [6, 23], the ground state neutron SF's for 61 nuclei have been calculated with uncertainties of about 10-20% [6]. Their predicted values are listed in Table 1 and plotted in the right panel of Figure 2. (In Figure 2, we exclude the deformed $^{24}$Mg, Li, F and Ne isotopes, some of which have small calculated or measured SF values, which, in general, tend not to be accurate.) In contrast to the IPM-plus-pairing calculations, the agreement between data and LB-SM predictions are within 20% for most cases, as indicated by the two dashed lines. For the $^{41-48}$Ca isotopes, there is close agreement between LB-SM predictions and IPM values as shown in Figures 1 and 2.

Due to the absorption of flux into other channels in the nuclear interior, the DWBA transfer integral samples the neutron bound state wavefunction mainly at the nuclear surface. Transfer reactions constrains the exterior but not the interior contributions to the overlap integral that defines SF. In this analysis, we assume a smooth variation of the potential radius ($R=r_0A^{1/3}$) for the bound neutron global potential. Consistent with findings in [24], we find the surface properties of the neutron bound-state wavefunction to be dominated by the central potential; for simplicity, we have neglected the spin-orbit interaction in constructing the neutron wavefunction. The good overall agreement between calculated and measured SF's indicates that these assumptions are reasonable, and that the relative magnitudes of SF's from nucleus to nucleus appear to be well described. However, the absolute values of the surface contributions to the SF are influenced by these geometrical assumptions.

It has long been asserted that transfer reactions do not yield absolute SF values. Nonetheless, it is informative to compare the results from different reaction mechanisms.

The (e,e'p) reaction removes protons from the same orbit for $^{12}C$, $^{16}O$, and $^{40}Ca$ isotopes as (p,d) reactions remove neutrons (and (d,p) reactions add neutrons). Assuming isospin symmetry, one might, therefore, expect the SF's determined from (p,d), (d,p) and (e,e'p) reactions to be comparable for these N=Z nuclei. Because transfer reactions are insensitive to depletions of the orbital occupancy due to hard core interactions in the dense nuclear interior, we expect the SF's from the (e,e'p) reactions, which probe the interior, to be about 10-15% lower than the SF's extracted from transfer reactions. This is indeed the case in the newest analysis on the $^{12}C(e,e'p)^{11}B$ data [25, 26]. However, the proton SF's from the older analysis of (e,e'p) are about 35-40% [27] smaller than the corresponding neutron SF's we extract for $^{12}C$, $^{16}O$, and $^{40}Ca$ isotopes. Any discrepancies between SF's extracted from (e,e'p) reactions and transfer reactions are intriquing but they are better resolved by extracting proton SF's from (d,$^{3}$He), ($^{3}$He,d), (p,n) and (n,p) reactions, using a similar analysis approach to that described in the present work.

Comparisons of neutron transfer reactions with nucleon knockout reactions using radioactive beams can also be made [28,29]. Recent measurements of SF's from single-nucleon "knock-out" reactions with radioactive and stable nuclei show increasing quenching of the SF values with nucleon separation energy, $S_n$ [28, 29]. No such dependence has been reported for (e,e'p) reactions [27]. Within the experimental uncertainties, we do not see any systematic quenching of the SF's with increasing nucleon separation energy [20]. Further theoretical study to understand the reaction mechanisms is needed to resolve the differences between the SF's obtained in (e,e'p), nucleon transfer and nucleon knockout reactions.

The structures of rare isotopes such as neutron rich nuclei with small neutron separation energy are of interest in astrophysical studies. For the seven nuclei with $S_n$<4 MeV ($^{8}Li$, $^{9}Be$, $^{11}Be$, $^{12}B$, $^{15}C$, $^{16}N$, $^{19}O$), most of the experimental SF values are smaller than the predictions from LB-SM calculations. It could be that the CH89 [11] potential may not be appropriate for the description of the scattering of these weakly bound nuclei with diffuse surfaces. Nevertheless, our present study of these nuclei provides reference points for future study with improved theoretical inputs [17].

In summary, we have extracted ground state neutron spectroscopic factors from transfer reactions for 80 nuclei using a consistent set of global input parameters and a DWBA integral that includes the effect of deuteron breakup in the (d,p) and (p,d) reactions. We find excellent agreement with large-basis shell-model calculations for most of the isotopes, suggesting that the current approach can be applied readily to other nuclei [19,30]. The spectroscopic factors obtained over a wide range of nuclei provide important benchmarks for neutron-transfer reaction studies. If additional radioactive nuclei are studied via nucleon-transfer and nucleon-knockout reactions, the results of current study provide important reference points for the development of advanced models to understand the mechanisms of these reactions.

The authors would like to thank Profs. J. Tostevin and B. A. Brown for helping us with using the codes TWOFNR and Oxbash. This work was supported by the National Science Foundation under Grant No. PHY-01-10253 and by the Summer for Undergraduate Research Experience (SURE) program at the Chinese University of Hong Kong.

Table 1. Spectroscopic factors (SF) obtained from transfer reactions (A(p,d)B and B(d,p)A) and from the large-basis shell-model (LB-SM) code Oxbash.

| A | LB-SM | SF | A | LB-SM | SF |
|---|---|---|---|---|---|
| $^{6}$Li | 0.68 | 1.08±0.22 | $^{34}$S | 1.83 | 1.38±0.20 |
| $^{7}$Li | 0.63 | 1.82±0.15 | $^{35}$S | 0.36 | 0.29±0.06 |
| $^{8}$Li | 1.09 | 0.61±0.12 | $^{37}$S | | 0.88±0.12 |
| $^{9}$Li | 0.81 | 0.98±0.09 | $^{35}$Cl | 0.32 | 0.33±0.07 |
| $^{9}$Be | 0.57 | 0.44±0.03 | $^{36}$Cl | 0.77 | 0.69±0.14 |
| $^{10}$Be | 2.36 | 1.53±0.15 | $^{37}$Cl | 1.15 | 1.03±0.15 |
| $^{11}$Be | 0.74 | 0.52±0.06 | $^{38}$Cl | | 1.74±0.35 |
| $^{10}$B | 0.60 | 0.49±0.07 | $^{36}$Ar | 2.06 | 3.23±0.46 |
| $^{11}$B | 1.09 | 1.34±0.12 | $^{37}$Ar | 0.36 | 0.35±0.04 |
| $^{12}$B | 0.83 | 0.45±0.06 | $^{38}$Ar | 3.04 | 2.43±0.49 |
| $^{12}$C | 2.85 | 2.98±0.30 | $^{39}$Ar | | 0.79±0.11 |
| $^{13}$C | 0.61 | 0.79±0.04 | $^{40}$Ar | | 1.05±0.21 |
| $^{14}$C | 1.73 | 1.56±0.13 | $^{41}$Ar | | 0.55±0.08 |
| $^{15}$C | 0.98 | 1.11±0.22 | $^{39}$K | 1.72 | 2.10±0.59 |
| $^{14}$N | 0.69 | 0.73±0.08 | $^{40}$K | | 1.66±0.33 |
| $^{15}$N | 1.46 | 1.38±0.11 | $^{41}$K | | 0.95±0.19 |
| $^{16}$N | 0.96 | 0.42±0.08 | $^{42}$K | | 0.77±0.11 |
| $^{16}$O | 2.00 | 2.23±0.13 | $^{40}$Ca | 4.00 | 4.30±0.38 |
| $^{17}$O | 1.00 | 0.84±0.04 | $^{41}$Ca | 1.00 | 0.99±0.05 |
| $^{18}$O | 1.58 | 1.75±0.20 | $^{42}$Ca | 1.81 | 1.97±0.18 |
| $^{19}$O | 0.69 | 0.41±0.06 | $^{43}$Ca | 0.75 | 0.62±0.07 |
| $^{19}$F | 0.56 | 1.56±0.22 | $^{44}$Ca | 3.64 | 4.37±0.50 |
| $^{20}$F | 0.02 | 0.01±0.00 | $^{45}$Ca | 0.50 | 0.39±0.06 |
| $^{21}$Ne | 0.03 | 0.03±0.00 | $^{47}$Ca | 0.26 | 0.25±0.04 |
| $^{22}$Ne | 0.01 | 0.23±0.03 | $^{48}$Ca | 7.38 | 7.05±0.81 |
| $^{23}$Ne | 0.03 | 0.24±0.03 | $^{49}$Ca | 0.92 | 0.68±0.07 |
| $^{24}$Na | 0.39 | 0.56±0.11 | $^{45}$Sc | 0.35 | 0.29±0.06 |
| $^{24}$Mg | 0.22 | 0.42±0.08 | $^{46}$Sc | 0.37 | 0.49±0.10 |
| $^{25}$Mg | 0.34 | 0.29±0.03 | $^{46}$Ti | 2.58 | 2.38±0.34 |
| $^{26}$Mg | 2.51 | 2.79±0.23 | $^{47}$Ti | | 0.01±0.00 |
| $^{27}$Mg | 0.46 | 0.44±0.09 | $^{48}$Ti | | 0.12±0.01 |
| $^{27}$Al | 1.10 | 1.35±0.19 | $^{49}$Ti | | 0.23±0.03 |
| $^{28}$Al | 0.60 | 0.64±0.09 | $^{50}$Ti | | 6.25±0.63 |
| $^{28}$Si | 3.62 | 4.23±0.85 | $^{51}$Ti | | 1.21±0.24 |
| $^{29}$Si | 0.45 | 0.37±0.03 | $^{51}$V | | 1.49±0.17 |
| $^{30}$Si | 0.82 | 0.72±0.10 | $^{50}$Cr | | 0.11±0.03 |
| $^{31}$Si | 0.58 | 0.59±0.07 | $^{51}$Cr | | 0.27±0.04 |
| $^{32}$P | 0.60 | 0.53±0.07 | $^{52}$Cr | | 6.03±0.85 |
| $^{32}$S | 0.96 | 1.46±0.41 | $^{53}$Cr | | 0.38±0.03 |
| $^{33}$S | 0.61 | 0.67±0.13 | $^{55}$Cr | | 0.86±0.17 |

Table II: List of nuclei with spectroscopic factors obtained from both (p,d) and (d,p) reactions. $N_{pd}$ and $N_{dp}$ denote the number of (p,d) and (d,p) independent measurements included in the analysis.

| A | A(p,d)B | $N_{pd}$ | B(d,p)A | $N_{dp}$ |
|---|---|---|---|---|
| $^{11}B$ | 1.25 | 2 | 1.44 | 3 |
| $^{11}Be$ | 0.56 | 1 | 0.46 | 2 |
| $^{13}C$ | 0.83 | 4 | 0.71 | 13 |
| $^{14}C$ | 1.30 | 4 | 1.75 | 2 |
| $^{15}N$ | 1.41 | 2 | 1.33 | 4 |
| $^{17}O$ | 0.77 | 4 | 0.95 | 10 |
| $^{18}O$ | 1.68 | 2 | 1.80 | 1 |
| $^{21}Ne$ | 0.03 | 1 | 0.03 | 2 |
| $^{26}Mg$ | 3.07 | 3 | 2.51 | 3 |
| $^{30}Si$ | 0.82 | 1 | 0.62 | 1 |
| $^{42}Ca$ | 2.14 | 2 | 1.77 | 3 |
| $^{43}Ca$ | 0.64 | 1 | 0.62 | 2 |
| $^{44}Ca$ | 4.16 | 3 | 5.00 | 1 |
| $^{48}Ti$ | 0.11 | 5 | 0.13 | 1 |
| $^{49}Ti$ | 0.24 | 2 | 0.23 | 1 |
| $^{50}Ti$ | 5.14 | 2 | 7.36 | 2 |
| $^{51}V$ | 1.61 | 1 | 1.31 | 2 |
| $^{53}Cr$ | 0.37 | 1 | 0.39 | 8 |

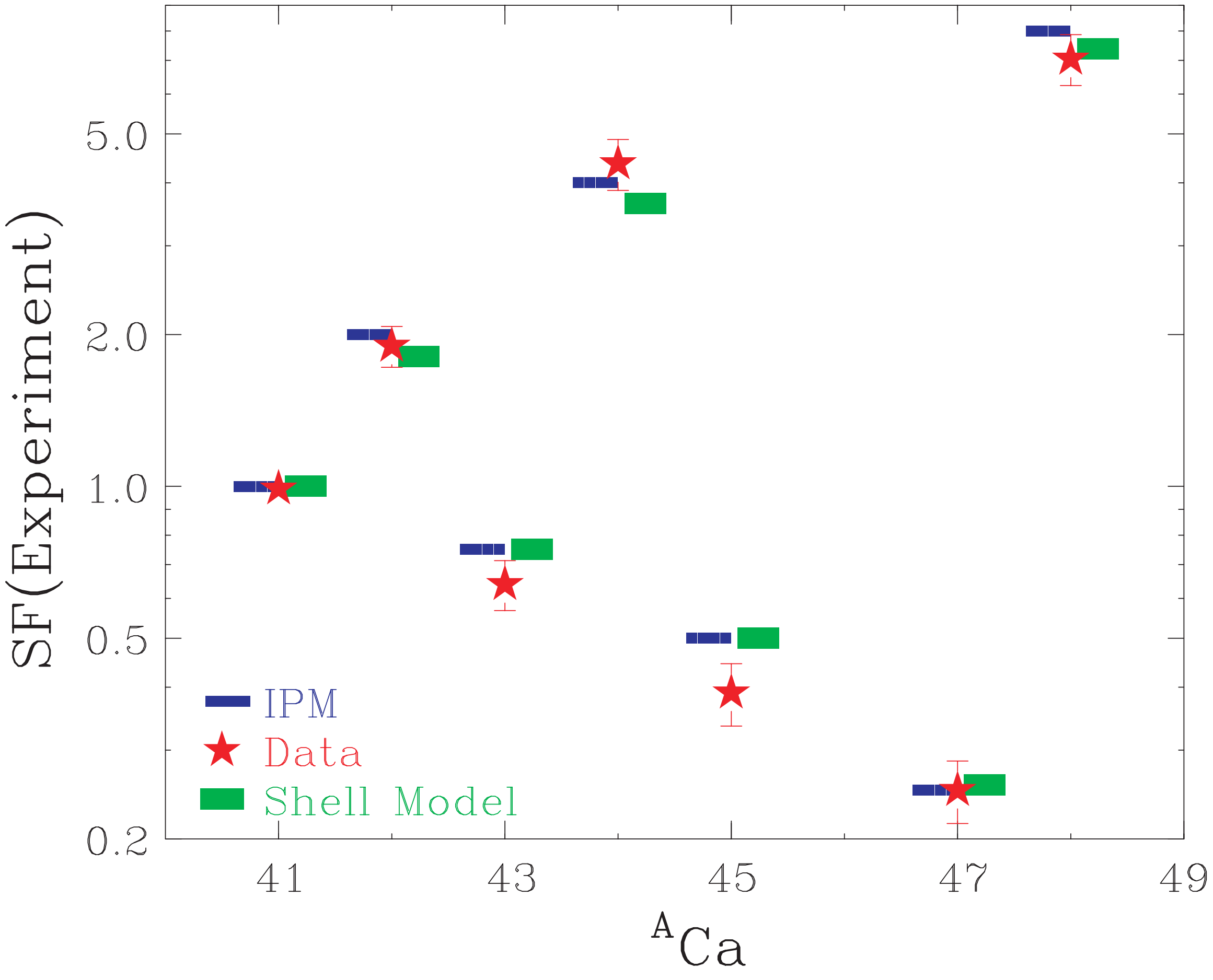

Figure 1 : (Color online) Ground state neutron spectroscopic factors for Calcium isotopes with valence neutrons in the $f_{7/2}$ orbit, star symbols represent SF values extracted from present analysis. Thin bars are IPM values and thick bars represent predictions from LB-SM using the program Oxbash.

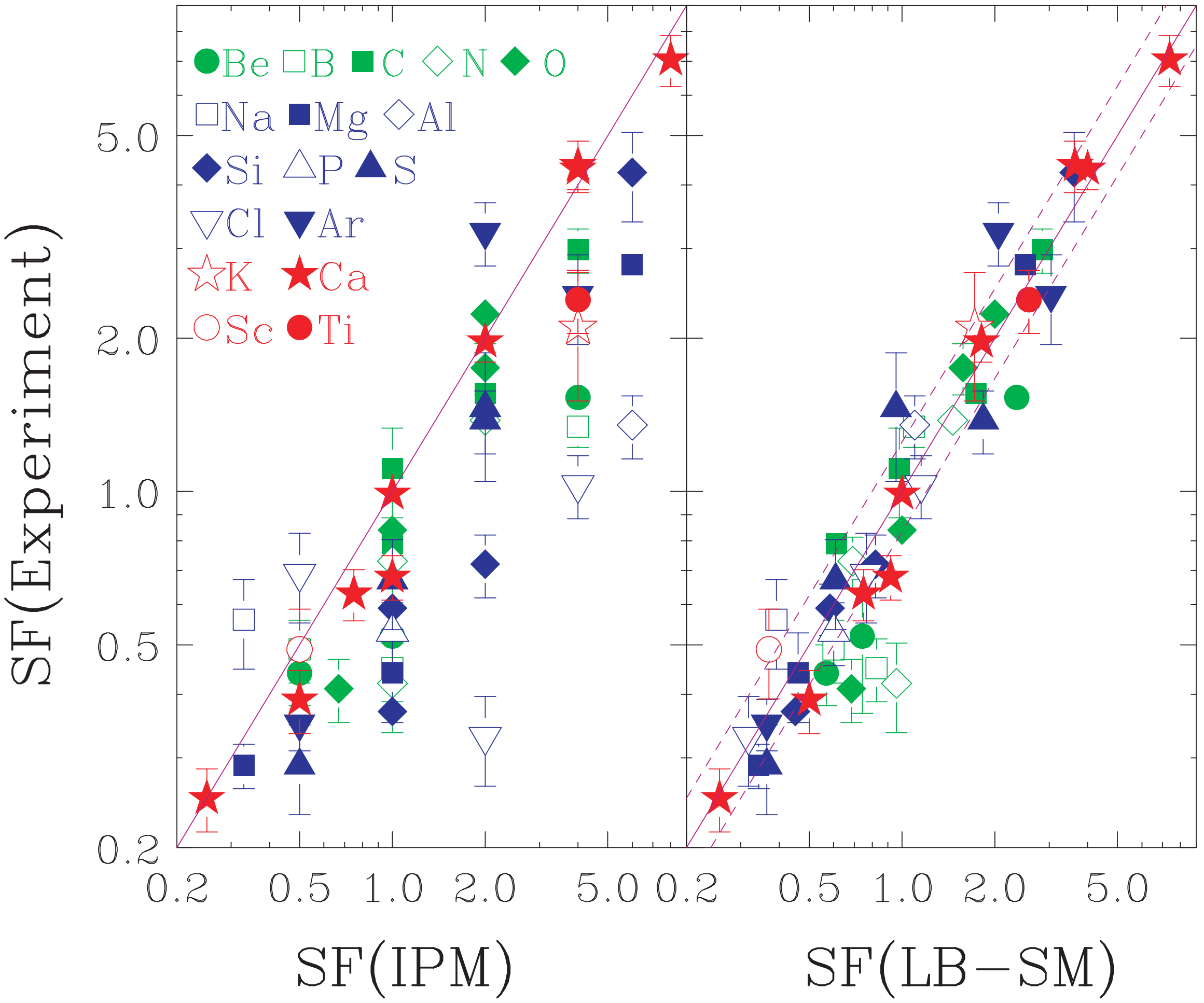

**Figure 2 :** (Color online) Comparison of experimental spectroscopic factors to predictions from the independent particle model of Eq. (1) (left panel) and LB-SM (right panel). Open and closed symbols denote elements with odd and even Z respectively. The three different colors of green, blue and red represent Z=3-8, 9-18 and 19-22 isotopes respectively. The solid lines indicate perfect agreement. For reference, the two dashed lines in the right panel indicate ±20% of the solid line.